\documentclass[sigconf,9pt,final]{acmart}

\usepackage{tikz}
\usepackage{tikz-layers}
\usepackage{subcaption}

\usetikzlibrary{arrows.meta}
\usetikzlibrary{positioning}



\title{Rearchitecting Kubernetes for the Edge}

\author{Andrew Jeffery}
\orcid{0000-0003-0440-0493}
\affiliation{%
	\institution{University of Cambridge}
	\department{Department of Computer Science and Technology}
	\city{Cambridge}
	\country{United Kingdom}
}
\email{andrew.jeffery@cl.cam.ac.uk}

\author{Heidi Howard}
\orcid{0000-0001-5256-7664}
\affiliation{%
	\institution{University of Cambridge}
	\department{Department of Computer Science and Technology}
	\city{Cambridge}
	\country{United Kingdom}
}
\email{heidi.howard@cl.cam.ac.uk}

\author{Richard Mortier}
\orcid{0000-0001-5205-5992}
\affiliation{%
	\institution{University of Cambridge}
	\department{Department of Computer Science and Technology}
	\city{Cambridge}
	\country{United Kingdom}
}
\email{richard.mortier@cl.cam.ac.uk}

\begin{abstract}

	Recent years have seen Kubernetes emerge as a primary choice for container orchestration.
	Kubernetes largely targets the cloud environment but new use cases require performant, available and scalable orchestration at the edge.
	Kubernetes stores all cluster state in \emph{etcd}, a strongly consistent key-value store.
	We find that at larger \emph{etcd} cluster sizes, offering higher availability, write request latency significantly increases and throughput decreases similarly.
	Coupled with approximately 30\% of Kubernetes requests being writes, this directly impacts the request latency and availability of Kubernetes, reducing its suitability for the edge.
	We revisit the requirement of strong consistency and propose an eventually consistent approach instead.
	This enables higher performance, availability and scalability whilst still supporting the broad needs of Kubernetes.
	This aims to make Kubernetes much more suitable for performance-critical, dynamically-scaled edge solutions.

\end{abstract}

\acmYear{2021}\copyrightyear{2021}
\setcopyright{rightsretained}
\acmConference[EdgeSys '21]{4th International Workshop on Edge Systems, Analytics and Networking}{April 26, 2021}{Online, United Kingdom}
\acmBooktitle{4th International Workshop on Edge Systems, Analytics and Networking (EdgeSys '21), April 26, 2021, Online, United Kingdom}
\acmPrice{}
\acmDOI{10.1145/3434770.3459730}
\acmISBN{978-1-4503-8291-5/21/04}

\begin{document}

\maketitle

\section{Introduction} \label{sec:introduction}

Recent years have seen containerisation and the associated orchestration become widespread in industry.
Kubernetes~\cite{Kubernetes}, a container orchestration platform, has emerged as a prominent solution in datacenters.
Edge use cases, with many thousands of nodes with limited CPU cores and RAM, are now becoming more prevalent, presenting the need for performant, available and reliable orchestration at the edge.

Kubernetes has been largely adopted across industry, with 59\% of large organisations using it in production~\cite{Tanzu}.
Kubernetes' flexibility can enable Functions as a Service, storage orchestration~\cite{Rook} and public-cloud integrations~\cite{CloudController}, and more.
This adoption and flexibility make Kubernetes an attractive platform for edge deployments.
Kubernetes uses \emph{etcd}~\cite{Etcd}, a strongly consistent distributed key-value store, as a source of truth for all control-plane components.
This makes \emph{etcd} a key factor in the path of all requests.
A background on Kubernetes and \emph{etcd} is provided in \textsection\ref{sec:background}.

While attractive, deploying Kubernetes at the edge still poses some challenges.
Both Kubernetes and \emph{etcd} can be resource intensive~\cite{KubernetesHardware,EtcdHardware}, leading to dedicated efforts to target Kubernetes towards the edge~\cite{K3s, K0s, KubeEdge}.
Edge environments typically have lower bandwidth, higher latency network connections, especially to non-local services than cloud datacenters.
Contrastingly these environments may be spread over much vaster scales and are expected to be more responsive to user interactions due to proximity whilst tolerating multiple failure classes.
With these harsh conditions performant, reliable and scalable orchestration is key.
We investigate the performance limitations of \emph{etcd}, their impact on scalability and on Kubernetes in \textsection\ref{sec:results}.

Kubernetes' reliance on \emph{etcd} and its limited scalability lead to both availability issues as well as efficiency issues at the edge.
Ultimately Kubernetes is limited by a fundamental design decision: the reliance on strong consistency in the datastore.
By revisiting this design decision in \textsection\ref{sec:solution} we aim to enable more performant, available and scalable orchestration at the edge.

Without a reliance on strong consistency architectural changes can become easier, especially with a focus on performance, availability and scalability.
The implications along with related work are discussed in \textsection\ref{sec:implications} and \textsection\ref{sec:related}.

The key contributions of this paper are:
\begin{enumerate}
	\item an explanation of how \emph{etcd} can be a bottleneck in a Kubernetes cluster, \textsection\ref{sec:background}
	\item a benchmark of \emph{etcd}'s performance at scale and discussion of the impact on availability, \textsection\ref{sec:results}
	\item revisiting the design decision of strong consistency and proposing to use eventual consistency, \textsection\ref{sec:solution}
\end{enumerate}

\section{Kubernetes and etcd} \label{sec:background}

Kubernetes organises containers into groups called Pods.
Pods are assigned to worker nodes where a local daemon (the \emph{Kubelet}) manages their lifecycle.
Higher level resources are used to implement concepts such as replicated Pods and services in the control-plane.
This control-plane is composed of Pods which reside on leader nodes, implementing core functionality such as the scheduler and API server.
These control-plane components are stateless and scale horizontally to aid performance and redundancy.
The desired cluster state and current status of applications, nodes and other resources is stored in an \emph{etcd} cluster.

Kubernetes is typically deployed in datacenter environments, typified by high bandwidth, low latency network connections.
Ideally, deployments should be spread across multiple datacenters for high availability~\cite{EtcdHardware}.
This requires that leader nodes run in each datacenter along with \emph{etcd} nodes to tolerate limited failures.
However, deploying \emph{etcd} across datacenters highlights the trade-off between availability and consistency \emph{etcd} faces as it scales.
This arises from the CAP theorem~\cite{Fox1999}, with \emph{etcd} sacrificing availability during network partitions to retain strong consistency.

\subsection{Etcd}

\emph{Etcd} is a strongly consistent, distributed, key-value store which
uses the \emph{Raft} consensus protocol~\cite{Ongaro2019} to maintain consistency, requiring a majority quorum.
As a component of the Kubernetes control-plane it can be deployed on standalone machines or hosted inside the Kubernetes cluster like any other control-plane service.
However, \emph{etcd} is not horizontally scalable due to overheads of maintaining strong consistency with more nodes.
Deployment recommendations suggest \emph{etcd} clusters should be set up with 3 or 5 nodes to attain high availability while avoiding these overheads~\cite{EtcdScale}.

\begin{table}
	\centering
	\caption{\emph{Etcd} request counts. Range requests are all linearisable. Requests with negligible count are omitted.}
	\begin{tabular}{lrr}
		\toprule
		\textbf{Request type} & \textbf{Count} & \textbf{Percentage} \\
		\midrule
		Range                 & 1542           & 52.3                \\
		Txn Range             & 476            & 16.1                \\
		Txn Put               & 866            & 29.3                \\
		Watch create          & 67             & 2.3                 \\
		\midrule
		Total                 & 2951           & 100                 \\
		\bottomrule
	\end{tabular}
	\label{tab:breakdown}
\end{table}

Table~\ref{tab:breakdown} contains a breakdown of the Kubernetes requests to a single node \emph{etcd} cluster for some basic Kubernetes operations, including setup, averaged over 10 runs.
The operations carried out were creating a deployment of 3 Pods, scaling it up to 10, back down to 5 and then deleting the deployment.
This provides a very simple trace of requests for scaling service deployments in Kubernetes.
\emph{Range} requests are \emph{get}s over multiple keys, \emph{put}s are \emph{write}s to a single key, these requests can also be contained within a transaction (\emph{txn}).
A \emph{watch create} request tells \emph{etcd} to notify the requester of changes to any keys in the provided range.
From this table we can see that \emph{put}s make up approximately 30\% of the total requests.
\emph{Put} requests may increase in proportion over the cluster lifetime as changes become more frequent and components rely on watches for updates rather than polling with \emph{range} requests.
\emph{Etcd}'s efficient handling of writes is therefore an important factor for Kubernetes.

\subsection{Scheduling walkthrough} \label{sec:walkthrough}

\begin{figure}
	\begin{tikzpicture}[every text node part/.style={align=center}]
		\node[draw,fill=green!20] (etcd1) {Etcd node 1\\(leader)};
		\node[draw,fill=green!20] (etcd2) [below left=1cm and 0cm of etcd1] {Etcd node 2};
		\node[draw,fill=green!20] (etcd3) [below right=1cm and 0cm of etcd1] {Etcd node 3};

		\node[draw,fill=red!20] (api) [above=1cm of etcd1, minimum width=5cm] {API server};

		\node[draw,fill=red!20] (replicaset) [above left=1cm and -1cm of api] {Replicaset\\controller};
		\node[draw,fill=red!20] (scheduler) [above right=1cm and -2.5cm of api] {Scheduler};
		\node[draw,fill=blue!20] (kubelet) [above right=1cm and 0cm of api] {Kubelet};
		\node[draw,fill=yellow!20] (registry) [above=of kubelet] {Container\\registry};
		\node[draw,fill=red!20] (custom) [above left=1cm and -1cm of scheduler] {Custom controllers /\\Deployment controller};

		\path[draw,-Latex] (api.173) -- (replicaset.250) node [midway, fill=white, draw] {2};
		\path[draw,-Latex] (replicaset) -- (api.170) node [midway, fill=white, draw] {3};
		\path[draw,-Latex] (api) -- (scheduler.210) node [midway, fill=white, draw] {7};
		\path[draw,-Latex] (scheduler.330) -- (api.20) node [midway, fill=white, draw] {8};
		\path[draw,-Latex] (api.10) -- (kubelet.210) node [midway, fill=white, draw] {12};
		\path[draw,Latex-Latex] (kubelet) -- (registry) node [midway, fill=white, draw] {13};
		\path[draw,-Latex] (kubelet.330) -- (api.7) node [midway, fill=white, draw] {14};

		\path[draw,Latex-Latex] (custom) -- (api.160) node [midway, fill=white!50, draw] {$\star$};
		\path[draw,-Latex] (api.345) -- (etcd1.35) node [midway, fill=white, draw] {4, 9, 15, $\star$};
		\path[draw,-Latex] (etcd1.150) -- (api.190) node [midway, fill=white, draw] {1, 6, 11, $\star$};
		\path[draw,Latex-Latex] (etcd1) -- (etcd2) node [midway, fill=white, draw=red, thick] {5, 10, 16, $\star$};
		\path[draw,Latex-Latex] (etcd1) -- (etcd3) node [midway, fill=white, draw=red, thick] {5, 10, 16, $\star$};
	\end{tikzpicture}

	\Description{Graph showing the route of requests to the \emph{etcd} cluster.}
	\caption{Flow of requests to schedule a Pod. Control-plane components are in
		red, \emph{etcd} nodes in green, node-local components in blue and cluster-external
		components in yellow.}

	\label{fig:schedvisual}
\end{figure}
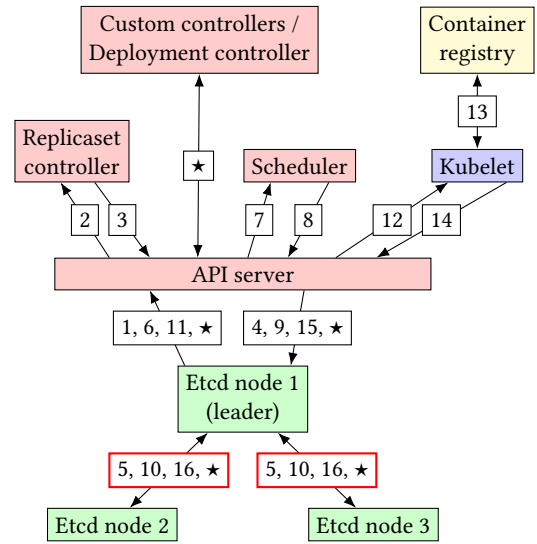

This section walks through the steps required to schedule a new Pod as part of a \emph{ReplicaSet}.
The steps are visualised in Figure~\ref{fig:schedvisual}.
Scheduling a new Pod can be a typical part of the process of reacting to a change of the \emph{replicas} field on a \emph{ReplicaSet} resource.
This change of value could originate from a failed node, an autoscaler or a manual scaling.

Step 1 and 2 see the updated \emph{ReplicaSet} resource being sent to the \emph{ReplicaSet} controller due to its existing \emph{watch}.
This controller then determines the necessary actions, creating a new \emph{Pod} resource in this case.
Steps 3 and 4 see this \emph{Pod} resource being written back to \emph{etcd}.
Due to the strong consistency of \emph{etcd}, step 5 is required to reach a majority quorum for the write.

Steps 6 and 7 see the new \emph{Pod} resource get passed to the scheduler.
This is also from a registered \emph{watch}, but this time on \emph{Pod} resources.
The \emph{Pod} resource does not currently specify the node to run on.
The scheduler filters suitable nodes down and selects an appropriate one to run the Pod.
The scheduler then writes the updated \emph{Pod} resource, with an assigned node, back to \emph{etcd} in steps 8 and 9.
Again, this write needs to be propagated to a majority quorum in step 10.

With an updated \emph{Pod} resource which has an associated node the \emph{Kubelet} gets notified of the update in steps 11 and 12.
With this complete \emph{Pod} description the \emph{Kubelet} begins the setup process for the containers.
This includes pulling the container images from a container registry in step 13.
During the setup process of the \emph{Pod}, events will be written to the resource in \emph{etcd}.
This occurs in steps 14 and 15 with the associated \emph{etcd} majority quorum writes in step 16.

After these steps the Pod should be set up and running on the node, managed by the \emph{Kubelet}.
More events will continue to happen such as the \emph{ReplicaSet} resource being updated with the new replica count.
It is also worth noting that a \emph{ReplicaSet} resource is typically controlled by a \emph{Deployment} resource, adding extra layers of communication and latency.
These added layers can be extended further due to Kubernetes custom controllers and resources, leading to significantly increased communication and scheduling latency along with a later initialisation of the Pod.
These are represented by a $\star$ in Figure~\ref{fig:schedvisual}.

As can be seen, there are lots of steps, each requiring separate writes to \emph{etcd} and thus quorum of the cluster.
Each of these increases the latency for scheduling a Pod and becomes a part of the dependency chain impacting reliability and availability.
While quorum writes within the \emph{etcd} cluster are sent in parallel the overall latency is dictated by the slowest node in the quorum.
In particular, this situation is exacerbated with a large cluster due to more load on the leader for communication and nodes being less likely to all operate within the same bounds.

\section{Etcd performance} \label{sec:results}

Operating \emph{etcd} for performance and availability in challenging large environments, such as the edge, requires it to scale efficiently while retaining performance.
This section outlines some initial results of testing \emph{etcd}'s scalability.

The tests used the official \emph{etcd} benchmarking tool\footnote{\url{https://github.com/etcd-io/etcd/blob/master/Documentation/op-guide/performance.md\#benchmarks}} with two different operations: \emph{put} and \emph{linearisable range} requests.
In \emph{etcd}, linearisable reads must return the value reflecting the consensus of the cluster.
Each run used only a single request type.

For each run a number of \emph{etcd} nodes, at version 3.4.13, were instantiated in Docker containers and arranged into a cluster with secure communication over a Docker network.
Each container was limited to 2 CPUs and 1GB RAM, using an SSD for data storage.
The host machine was running Linux, kernel 4.15.0, on an Intel Xeon 4112, 16 core CPU with 196GB of RAM.
Each test configuration was repeated 10 times and medians of these repeats are presented.
The benchmark targeted all nodes, not just the leader, using 1,000 clients, each with 100 connections, performing 100,000 operations in total in each run.
This aims to provide a best case scenario for \emph{etcd}'s performance and scalability in an idealised setting without network latency.
Network interactions would add further variability and instability to the system, enabling more failure scenarios such as partial partitions~\cite{alfatafta2020toward, CloudFlare}.

\begin{figure}
	\begin{subfigure}[b]{\linewidth}
		\includegraphics[width=\linewidth]{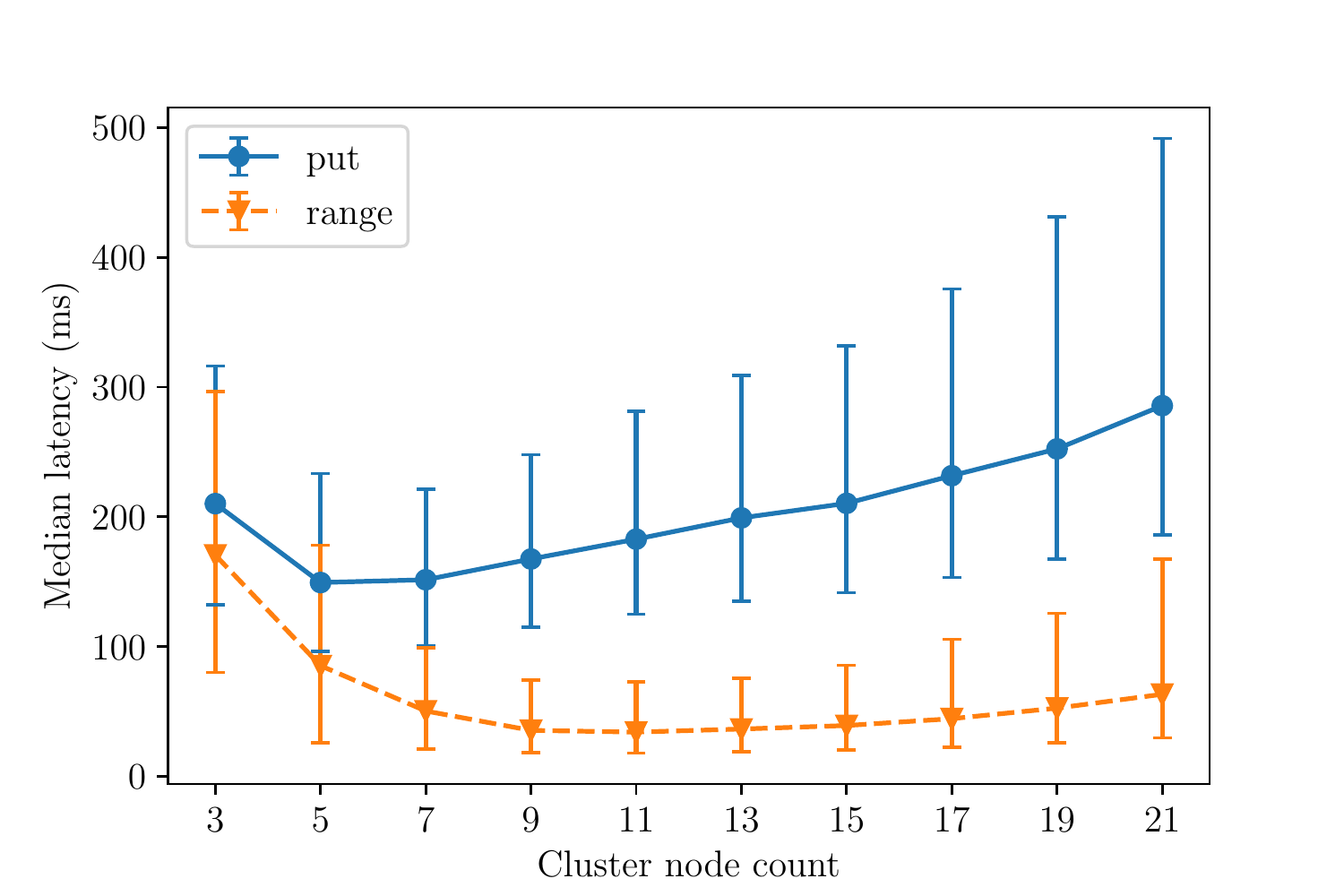}
		\Description{A line plot showing linearly-increasing \emph{put} latency with slightly increasing range latency.}
		\caption{Median latency, error bars at p10 and p90.}
		\label{fig:results-latency}
	\end{subfigure}
	\begin{subfigure}[b]{\linewidth}
		\includegraphics[width=\linewidth]{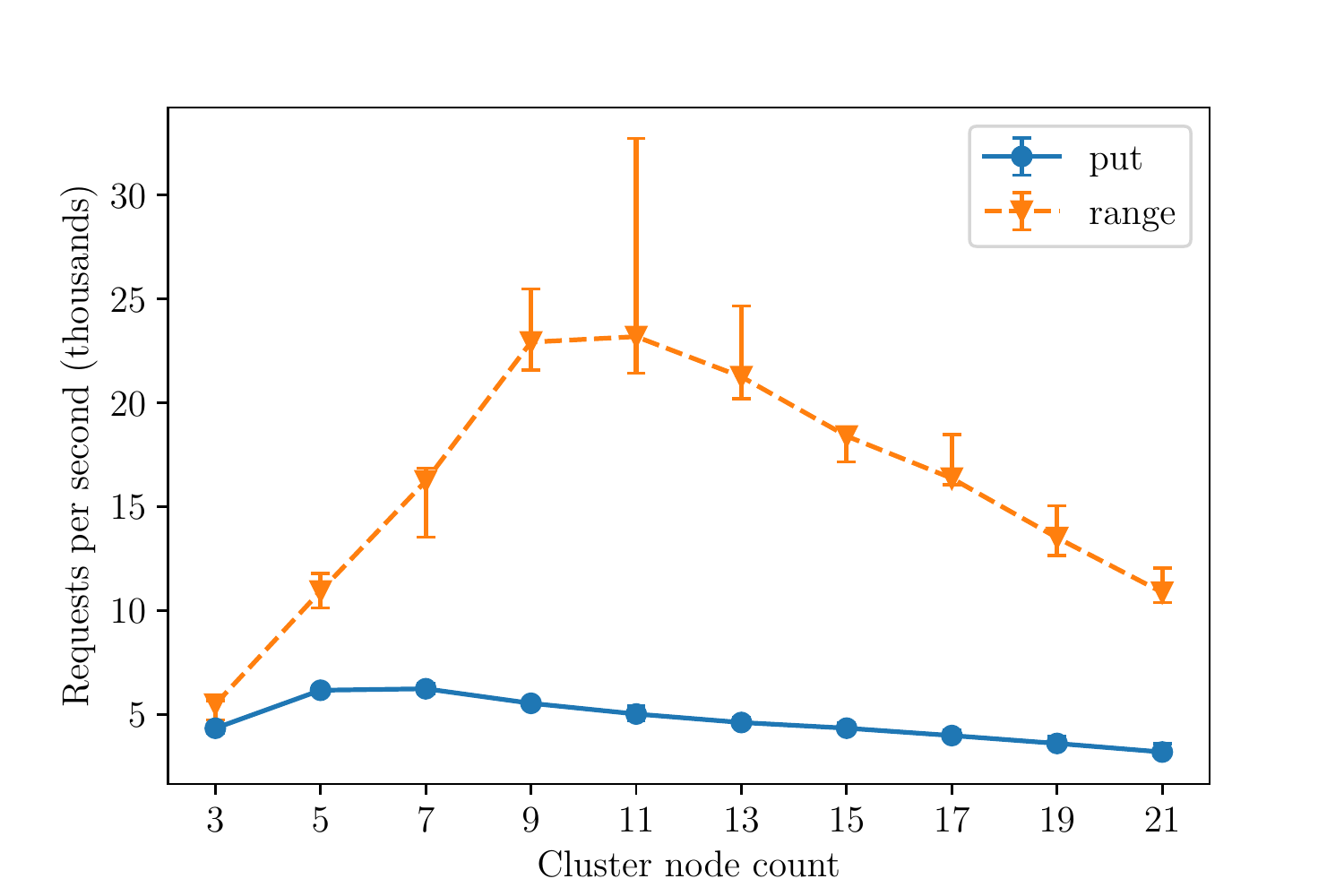}
		\Description{A line plot showing decreasing \emph{put} throughput with decreasing range throughput.}
		\caption{Median throughput, error bars at p10 and p90.}
		\label{fig:results-throughput}
	\end{subfigure}
	\caption{Results of scalability testing with \emph{etcd}.}
	\label{fig:results}
\end{figure}

Figure~\ref{fig:results-latency} shows that the strong consistency of writes certainly comes at a cost in terms of latency, having to write the value to a majority of nodes each time.
Meanwhile, the read latency stays comparatively low, avoiding the latency impact of flushing writes to disk.
As the cluster size increases the amount of synchronisation work done by a leader node increases, causing the observed decrease in performance.
With an eventually consistent datastore the latency of both reads and writes would be expected to remain similar to each other and decrease as the cluster scales by spreading the load more efficiently.

Figure~\ref{fig:results-throughput} shows the effect of increasing node counts on throughput.
For both scenarios, large \emph{etcd} cluster sizes lead to a severe degradation of throughput, regardless of request type.
The requests require a majority quorum leading to lots of inter-node requests, ultimately being a bottleneck and lowering throughput.
Running an eventually consistent datastore would lead to throughput increasing with scale as there is no coordination during the request, similar to the results observed in Anna~\cite{Anna}.

Due to this limited performance at scale, \emph{etcd} imposes a trade-off of performance or availability.
This limit on availability can leave Kubernetes clusters unable to make progress in the event of failures or scale services to cope with demand.
These results, coupled with \emph{put}s forming a significant proportion of Kubernetes requests, show that \emph{etcd} and such strongly consistent datastores are not going to be sufficient in the harsher conditions of the edge environment.

\section{Eventually consistent datastore} \label{sec:solution}

This section outlines the planned work to replace \emph{etcd} with an eventually consistent datastore and some implementation considerations.

\begin{figure}
	\begin{tikzpicture}[every text node part/.style={align=center}]
		\node[draw,fill=green!20] (etcd1) {Datastore\\node 1};
		\node[draw,fill=green!20] (etcd2) [below left=1cm and 0.5cm of etcd1] {Datastore\\node 2};
		\node[draw,fill=green!20] (etcd3) [below right=1cm and 0.5cm of etcd1] {Datastore\\node 3};
		\node[draw,fill=green!20] (etcdmany) [below =1cm of etcd1] {$\cdots$};

		\node[draw,fill=red!20] (api) [above=1cm of etcd1, minimum width=5cm] {API server};

		\node[draw,fill=red!20] (replicaset) [above left=1cm and -1cm of api] {Replicaset\\controller};
		\node[draw,fill=red!20] (scheduler) [above right=1cm and -2.5cm of api] {Scheduler};
		\node[draw,fill=blue!20] (kubelet) [above right=1cm and 0cm of api] {Kubelet};
		\node[draw,fill=yellow!20] (registry) [above=of kubelet] {Container\\registry};
		\node[draw,fill=red!20] (custom) [above left=1cm and -1cm of scheduler] {Custom controllers /\\Deployment controller};

		\path[draw,-Latex] (api.173) -- (replicaset.250) node [midway, fill=white, draw] {2};
		\path[draw,-Latex] (replicaset) -- (api.170) node [midway, fill=white, draw] {3};
		\path[draw,-Latex] (api) -- (scheduler.210) node [midway, fill=white, draw] {6};
		\path[draw,-Latex] (scheduler.330) -- (api.20) node [midway, fill=white, draw] {7};
		\path[draw,-Latex] (api.10) -- (kubelet.210) node [midway, fill=white, draw] {10};
		\path[draw,-Latex] (kubelet) -- (registry) node [midway, fill=white, draw] {11};
		\path[draw,-Latex] (kubelet.330) -- (api.7) node [midway, fill=white, draw] {12};

		\path[draw,Latex-Latex] (custom) -- (api.160) node [midway, fill=white, draw] {$\star$};
		\path[draw,-Latex] (api.345) -- (etcd1.30) node [midway, fill=white, draw] {4, 8, 13, $\star$};
		\path[draw,-Latex] (etcd1.150) -- (api.190) node [midway, fill=white, draw] {1, 5, 9, $\star$};
		\path[draw,Latex-Latex,dashed] (etcd1) -- (etcd2) node [midway, fill=white, draw=green, thick] {lazy syncing};
		\path[draw,Latex-Latex,dashed] (etcd1) -- (etcd3) node [midway, fill=white, draw=green, thick] {lazy syncing};
	\end{tikzpicture}

	\Description{Graph showing the route of requests to the proposed datastore cluster.}
	\caption{Flow of requests to schedule a Pod with the proposed datastore. Syncing between datastore nodes is now lazy, not interfering with the critical path of the request.}

	\label{fig:schednew}
\end{figure}

\subsection{The \emph{etcd} API}

Due to the coupling between the API server and \emph{etcd} cluster the proposed work will need to implement and expose the same API, though the inner workings and guarantees will differ.
This ensures that no changes to Kubernetes components would be required.

Some behaviours of the API exposed by the proposed work will not correspond to that of \emph{etcd} due to the difference in architecture.
For instance, reporting which node is the leader is non-sensical in the proposed work, instead it will likely report each node as a leader.

\subsection{Lazy syncing}

To implement the functionality of this API and attain low latency, the proposed datastore needs to allow reads and writes to a single node to be performed without immediate communication with other nodes.
This enables the possibility of concurrent writes to different nodes, introducing conflicts in the stored data.
Conflict-free replicated datatypes~\cite{Preguica2018} (CRDTs) enable these conflicts to be resolved upon syncing with other nodes in a lazy, rather than eager, manner.
This will enable fast responses to the API server even at large scales, as demonstrated in Figure~\ref{fig:schednew}, due to no requests between the datastore nodes in the critical path.

CRDTs come in two main varieties: state-based and operation-based.
To synchronise two replicas state-based CRDTs transfer the entire local state for combination with the remote state.
In contrast, operation-based CRDTs transfer operations to be applied on the remote state.
Operation-based CRDTs have minimal bandwidth requirements compared to state-based CRDTs though recent work has helped to close this gap~\cite{Almeida2015, VanderLinde, Enes2019}.

Kubernetes uses protobuf schema files to declare the format of resources to be stored in \emph{etcd}, resembling JSON.
These resources are not already CRDTs so this translation will be within the datastore.
This may require calculating the change between the new and stored values, extracting the operations to apply to the CRDT.
With these operations and the knowledge of the data format we can use a JSON CRDT~\cite{Kleppmann2017} to provide eventual consistency for Kubernetes resource objects.
Recent work~\cite{kleppmann2020byzantine} has also introduced low latency, single round trip syncing of operations between nodes in untrusted environments.
This can be applied to CRDTs providing an efficient method of synchronisation for the edge.

\subsection{Impact on Kubernetes}

From Table~\ref{tab:breakdown} we saw that transactions make up a large component of the requests to \emph{etcd}.
Due to a lack of consensus, transactions for the proposed datastore would only operate on data in the targeted store at the execution time.
This means that they could act on stale data with respect to other nodes.
However, probabilistically bounded staleness ~\cite{Bailis2150} shows that an eventually consistent system can often still present the latest updates to data.
Additionally, due to the control loop employed by Kubernetes components, any errors should be rectified quickly.
For instance, if two separate nodes increase the count on a \emph{ReplicaSet} resource concurrently, two new Pods may be scheduled.
When these datastore nodes synchronise, these changes may get combined into a total increase of 2 replicas.
The Kubernetes controllers can then observe this new value and decide whether this can remain or it should be decreased.

\section{Implications for architectures} \label{sec:implications}

Due to the improved scalability and lack of consensus in the proposed datastore it would be possible to use autoscaling.
This would enable more optimal resource usage, reacting to demand.
If the datastore is hosted inside the Kubernetes cluster then the native horizonal autoscaler could be employed as a low complexity solution.
This is currently not practical with \emph{etcd} due to scaling limitations coupled with the more static nature of strongly consistent systems.

The proposed datastore, while enabling higher availability deployments through scalability, also enables a partitioned datacenter to remain operational.
Remaining able to respond to failures or changes in demand is a key operational benefit as system failures often cause complex problems~\cite{CloudFlare}.

In an edge environment, the proposed datastore could be spread across the Kubernetes cluster at greater scale.
This enables utilising the horizontal scalability of the stateless control-plane to lower latencies, in particular for scheduling.
\emph{Etcd} cannot be scaled to this extent, imposing a lower limit on request latencies.

With this scalability it could be feasible to deploy control-plane components with a datastore on each worker node, making Kubernetes decentralised.
The current Kubernetes scheduler could then be replaced with a local-first distributed scheduler, leveraging the vast literature surrounding distributed scheduling~\cite{Zaharia2010, Ren2015, Stankovic, Stankovic1985}.
This rearchitecting would mean that the scheduling process would not require any requests to leave the originating node, drastically reducing scheduling latency.
This could enable efficient reactive autoscaling and potentially native Functions as a Service.

\section{Related work} \label{sec:related}

New use cases for edge environments include 5G networking~\cite{Bonati2020}, in-network computing~\cite{Krol2019} and elastic CDNs~\cite{Kuenzer}.
These all require orchestrating lots of machines at the edge with emphasis on low latency and reliability.
Recent work has seen Kubernetes already become popular for this orchestration at the edge~\cite{Dupont, Chen}.
Kubernetes, with a more performant and available core, can fit the orchestration frameworks required for these use cases to offload work from the cloud, improve latency for requests and provide service-level adaptability.

Federated Kubernetes~\cite{KubeFed} distributes work between clusters, consisting of a host cluster that is responsible for distributing the work between member clusters.
This centralised approach has a similar downside to a large single cluster, leading to new research into a decentralised model of federation using CRDTs~\cite{Larsson2020}.
This separates cluster-local state from federation state, focusing on just the federation state.
Our work instead tackles the problem of cluster-local state.

DOCMA~\cite{Jimenez2019} is a new orchestrator for container based microservices.
This achieves a decentralised architecture enabling deployments with several thousands of nodes.
However, this lacks a significant number of features Kubernetes provides.
DOCMA shows that decentralised orchestration is highly scalable and provides significant redundancy.

Proposed Kubernetes architectures for edge environments vary but are all constrained by the centralised state in \emph{etcd}.
Some propose hosting the leader nodes and \emph{etcd} in a datacenter and only worker nodes at the edge~\cite{K3s}.
However, connections to the cloud can have high latencies and be unreliable meaning further engineering is required to have a robust edge~\cite{SuperEdge, OpenYurt}.
Others propose deploying everything to the edge, including the datastore~\cite{KubeEdge}, though resource limitations can make this less viable.

Software defined networking has seen lots of research around consistency of control-plane state~\cite{Feamstera, Sakic2017, Koponen}.
Concepts such as adaptive consistency~\cite{Sakic2017} and data-partitioning based on consistency requirements~\cite{Koponen} may prove useful to augment our datastore.
Alternatively, strongly consistent systems can avoid the need for strict majority quorums, leading to more scalable systems~\cite{Ailijiang2017, Rystsov2018}.
However, these all inherit the trade-off of latency and consistency.
Instead, we focus on minimising latency to offer performance and availability in the challenging edge environment.

\section{Conclusion} \label{sec:conclusion}

This paper has highlighted the extensive reliance of Kubernetes on \emph{etcd} and the factors leading to lower availability and a delay in scheduling.
We observed that \emph{etcd} poses a bottleneck in cluster scalability, with an impact on scheduling latency and availability of the whole system due to its limited performance at scale.
Our results support our key observation that reliance on strong consistency in the datastore limits the performance, availability and scalability of Kubernetes.
We propose to build a decentralised, eventually consistent store specialised to Kubernetes in order to combat these issues.
This redesign also leads to the opportunity to rearchitect Kubernetes for edge environments, offering increased performance, availability and scalability.
These improvements could lead to lower latency, larger scale deployments at the edge and hope to inform the future of orchestration platforms, targeting decentralised approaches for availability and performance.

\section*{Acknowledgements}

This work is funded in part by EPSRC EP/R03351X/1, EP/M02315X/1 and EP/T022493/1.

\bibliography{../default}


\begin{thebibliography}{42}

%%% the Web address.

\ifx \showCODEN    \undefined \def \showCODEN     #1{\unskip}     \fi
\ifx \showDOI      \undefined \def \showDOI       #1{#1}\fi
\ifx \showISBNx    \undefined \def \showISBNx     #1{\unskip}     \fi
\ifx \showISBNxiii \undefined \def \showISBNxiii  #1{\unskip}     \fi
\ifx \showISSN     \undefined \def \showISSN      #1{\unskip}     \fi
\ifx \showLCCN     \undefined \def \showLCCN      #1{\unskip}     \fi
\ifx \shownote     \undefined \def \shownote      #1{#1}          \fi
\ifx \showarticletitle \undefined \def \showarticletitle #1{#1}   \fi
\ifx \showURL      \undefined \def \showURL       {\relax}        \fi
% invisible to TeX
\providecommand\bibfield[2]{#2}
\providecommand\bibinfo[2]{#2}
\providecommand\natexlab[1]{#1}
\providecommand\showeprint[2][]{arXiv:#2}

\bibitem[\protect\citeauthoryear{??}{Clo}{2020}]%
        {CloudFlare}
 \bibinfo{year}{2020}\natexlab{}.
\newblock \bibinfo{booktitle}{\emph{{A Byzantine failure in the real world}}}.
\newblock
\urldef\tempurl%
\url{https://blog.cloudflare.com/a-byzantine-failure-in-the-real-world/}
\showURL{%
Retrieved January 13, 2021 from \tempurl}


\bibitem[\protect\citeauthoryear{??}{Ope}{2020}]%
        {OpenYurt}
 \bibinfo{year}{2020}\natexlab{}.
\newblock \bibinfo{booktitle}{\emph{{An open platform that extends upstream
  Kubernetes to Edge}}}.
\newblock
\urldef\tempurl%
\url{https://openyurt.io/en-us/index.html}
\showURL{%
Retrieved January 13, 2021 from \tempurl}


\bibitem[\protect\citeauthoryear{??}{K3s}{2020}]%
        {K3s}
 \bibinfo{year}{2020}\natexlab{}.
\newblock \bibinfo{booktitle}{\emph{{K3s: The certified Kubernetes distribution
  built for IoT {\&} Edge computing}}}.
\newblock
\urldef\tempurl%
\url{https://k3s.io/}
\showURL{%
Retrieved January 13, 2021 from \tempurl}


\bibitem[\protect\citeauthoryear{??}{Kub}{2020a}]%
        {KubeEdge}
 \bibinfo{year}{2020}\natexlab{a}.
\newblock \bibinfo{booktitle}{\emph{{KubeEdge An open platform to enable Edge
  computing}}}.
\newblock
\urldef\tempurl%
\url{https://kubeedge.io/en/}
\showURL{%
Retrieved January 13, 2021 from \tempurl}


\bibitem[\protect\citeauthoryear{??}{Kub}{2020b}]%
        {KubeFed}
 \bibinfo{year}{2020}\natexlab{b}.
\newblock \bibinfo{booktitle}{\emph{{KubeFed: Kubernetes Cluster Federation}}}.
\newblock
\urldef\tempurl%
\url{https://github.com/kubernetes-sigs/kubefed}
\showURL{%
Retrieved January 13, 2021 from \tempurl}


\bibitem[\protect\citeauthoryear{??}{Sup}{2020}]%
        {SuperEdge}
 \bibinfo{year}{2020}\natexlab{}.
\newblock \bibinfo{booktitle}{\emph{{SuperEdge: An edge-native container
  management system for edge computing}}}.
\newblock
\urldef\tempurl%
\url{https://github.com/superedge/superedge}
\showURL{%
Retrieved January 13, 2021 from \tempurl}


\bibitem[\protect\citeauthoryear{??}{Clo}{2021}]%
        {CloudController}
 \bibinfo{year}{2021}\natexlab{}.
\newblock \bibinfo{booktitle}{\emph{{Cloud Controller Manager}}}.
\newblock
\urldef\tempurl%
\url{https://kubernetes.io/docs/concepts/architecture/cloud-controller/}
\showURL{%
Retrieved February 09, 2021 from \tempurl}


\bibitem[\protect\citeauthoryear{??}{Etc}{2021a}]%
        {Etcd}
 \bibinfo{year}{2021}\natexlab{a}.
\newblock \bibinfo{booktitle}{\emph{{Etcd: A distributed, reliable key-value
  store for the most critical data of a distributed system}}}.
\newblock
\urldef\tempurl%
\url{https://etcd.io/}
\showURL{%
Retrieved February 09, 2021 from \tempurl}


\bibitem[\protect\citeauthoryear{??}{Etc}{2021b}]%
        {EtcdHardware}
 \bibinfo{year}{2021}\natexlab{b}.
\newblock \bibinfo{booktitle}{\emph{{Etcd: Hardware recommendations}}}.
\newblock
\urldef\tempurl%
\url{https://etcd.io/docs/v3.4.0/op-guide/hardware}
\showURL{%
Retrieved February 09, 2021 from \tempurl}


\bibitem[\protect\citeauthoryear{??}{K0s}{2021}]%
        {K0s}
 \bibinfo{year}{2021}\natexlab{}.
\newblock \bibinfo{booktitle}{\emph{{K0s: The Simple, Solid {\&} Certified
  Kubernetes Distribution}}}.
\newblock
\urldef\tempurl%
\url{https://k0sproject.io/}
\showURL{%
Retrieved January 13, 2021 from \tempurl}


\bibitem[\protect\citeauthoryear{??}{Kub}{2021a}]%
        {KubernetesHardware}
 \bibinfo{year}{2021}\natexlab{a}.
\newblock \bibinfo{booktitle}{\emph{{Kubernetes kubeadm resource
  requirements}}}.
\newblock
\urldef\tempurl%
\url{https://kubernetes.io/docs/setup/production-environment/tools/kubeadm/create-cluster-kubeadm/}
\showURL{%
Retrieved February 16, 2021 from \tempurl}


\bibitem[\protect\citeauthoryear{??}{Kub}{2021b}]%
        {Kubernetes}
 \bibinfo{year}{2021}\natexlab{b}.
\newblock \bibinfo{booktitle}{\emph{{Kubernetes: Production-Grade Container
  Orchestration}}}.
\newblock
\urldef\tempurl%
\url{https://kubernetes.io/}
\showURL{%
Retrieved February 09, 2021 from \tempurl}


\bibitem[\protect\citeauthoryear{??}{Roo}{2021}]%
        {Rook}
 \bibinfo{year}{2021}\natexlab{}.
\newblock \bibinfo{booktitle}{\emph{{Rook: Open-Source, Cloud-Native Storage
  for Kubernetes}}}.
\newblock
\urldef\tempurl%
\url{https://rook.io/}
\showURL{%
Retrieved February 09, 2021 from \tempurl}


\bibitem[\protect\citeauthoryear{??}{Etc}{2021c}]%
        {EtcdScale}
 \bibinfo{year}{2021}\natexlab{c}.
\newblock \bibinfo{booktitle}{\emph{{Scaling up etcd clusters}}}.
\newblock
\urldef\tempurl%
\url{https://kubernetes.io/docs/tasks/administer-cluster/configure-upgrade-etcd/#scaling-up-etcd-clusters}
\showURL{%
Retrieved February 09, 2021 from \tempurl}


\bibitem[\protect\citeauthoryear{??}{Tan}{2021}]%
        {Tanzu}
 \bibinfo{year}{2021}\natexlab{}.
\newblock \bibinfo{booktitle}{\emph{{Why Large Organizations Trust
  Kubernetes}}}.
\newblock
\urldef\tempurl%
\url{https://tanzu.vmware.com/content/blog/why-large-organizations-trust-kubernetes}
\showURL{%
Retrieved March 31, 2021 from \tempurl}


\bibitem[\protect\citeauthoryear{{Ailijiang}, {Charapko}, {Demirbas}, and
  {Kosar}}{{Ailijiang} et~al\mbox{.}}{2020}]%
        {Ailijiang2017}
\bibfield{author}{\bibinfo{person}{Ailidani {Ailijiang}},
  \bibinfo{person}{Aleksey {Charapko}}, \bibinfo{person}{Murat {Demirbas}},
  {and} \bibinfo{person}{Tevfik {Kosar}}.} \bibinfo{year}{2020}\natexlab{}.
\newblock \showarticletitle{WPaxos: Wide Area Network Flexible Consensus}.
\newblock \bibinfo{journal}{\emph{IEEE Transactions on Parallel and Distributed
  Systems}} \bibinfo{volume}{31}, \bibinfo{number}{1} (\bibinfo{year}{2020}),
  \bibinfo{pages}{211--223}.
\newblock
\urldef\tempurl%
\url{https://doi.org/10.1109/TPDS.2019.2929793}
\showDOI{\tempurl}


\bibitem[\protect\citeauthoryear{Alfatafta, Alkhatib, Alquraan, and
  Al-Kiswany}{Alfatafta et~al\mbox{.}}{2020}]%
        {alfatafta2020toward}
\bibfield{author}{\bibinfo{person}{Mohammed Alfatafta}, \bibinfo{person}{Basil
  Alkhatib}, \bibinfo{person}{Ahmed Alquraan}, {and} \bibinfo{person}{Samer
  Al-Kiswany}.} \bibinfo{year}{2020}\natexlab{}.
\newblock \showarticletitle{Toward a Generic Fault Tolerance Technique for
  Partial Network Partitioning}. In \bibinfo{booktitle}{\emph{Operating Systems
  Design and Implementation ({OSDI}) 2020}}.
\newblock


\bibitem[\protect\citeauthoryear{Almeida, Shoker, and Baquero}{Almeida
  et~al\mbox{.}}{2015}]%
        {Almeida2015}
\bibfield{author}{\bibinfo{person}{Paulo~S{\`{e}}rgio Almeida},
  \bibinfo{person}{Ali Shoker}, {and} \bibinfo{person}{Carlos Baquero}.}
  \bibinfo{year}{2015}\natexlab{}.
\newblock \bibinfo{title}{{Efficient state-based CRDTs by delta-mutation}}.
\newblock
\newblock
\urldef\tempurl%
\url{https://doi.org/10.1007/978-3-319-26850-7_5}
\showDOI{\tempurl}


\bibitem[\protect\citeauthoryear{Bailis, Venkataraman, Franklin, Hellerstein,
  and Stoica}{Bailis et~al\mbox{.}}{2012}]%
        {Bailis2150}
\bibfield{author}{\bibinfo{person}{Peter Bailis}, \bibinfo{person}{Shivaram
  Venkataraman}, \bibinfo{person}{Michael~J. Franklin},
  \bibinfo{person}{Joseph~M. Hellerstein}, {and} \bibinfo{person}{Ion Stoica}.}
  \bibinfo{year}{2012}\natexlab{}.
\newblock \showarticletitle{Probabilistically Bounded Staleness for Practical
  Partial Quorums}.
\newblock \bibinfo{journal}{\emph{Proceedings of the VLDB Endowment}}
  \bibinfo{volume}{5}, \bibinfo{number}{8} (\bibinfo{date}{April}
  \bibinfo{year}{2012}), \bibinfo{pages}{776–787}.
\newblock
\showISSN{2150-8097}
\urldef\tempurl%
\url{https://doi.org/10.14778/2212351.2212359}
\showDOI{\tempurl}


\bibitem[\protect\citeauthoryear{Bonati, Polese, D’Oro, Basagni, and
  Melodia}{Bonati et~al\mbox{.}}{2020}]%
        {Bonati2020}
\bibfield{author}{\bibinfo{person}{Leonardo Bonati}, \bibinfo{person}{Michele
  Polese}, \bibinfo{person}{Salvatore D’Oro}, \bibinfo{person}{Stefano
  Basagni}, {and} \bibinfo{person}{Tommaso Melodia}.}
  \bibinfo{year}{2020}\natexlab{}.
\newblock \showarticletitle{Open, Programmable, and Virtualized 5G Networks:
  State-of-the-Art and the Road Ahead}.
\newblock \bibinfo{journal}{\emph{Computer Networks}}  \bibinfo{volume}{182}
  (\bibinfo{year}{2020}), \bibinfo{pages}{107516}.
\newblock
\showISSN{1389-1286}
\urldef\tempurl%
\url{https://doi.org/10.1016/j.comnet.2020.107516}
\showDOI{\tempurl}


\bibitem[\protect\citeauthoryear{{Chen} and {Lin}}{{Chen} and {Lin}}{2019}]%
        {Chen}
\bibfield{author}{\bibinfo{person}{Hung-Li {Chen}} {and}
  \bibinfo{person}{Fuchun~J. {Lin}}.} \bibinfo{year}{2019}\natexlab{}.
\newblock \showarticletitle{Scalable IoT/M2M Platforms Based on
  Kubernetes-Enabled NFV MANO Architecture}. In
  \bibinfo{booktitle}{\emph{International Conference on Internet of Things
  (iThings) 2019}}.
\newblock
\urldef\tempurl%
\url{https://doi.org/10.1109/iThings/GreenCom/CPSCom/SmartData.2019.00188}
\showDOI{\tempurl}


\bibitem[\protect\citeauthoryear{{Dupont}, {Giaffreda}, and {Capra}}{{Dupont}
  et~al\mbox{.}}{2017}]%
        {Dupont}
\bibfield{author}{\bibinfo{person}{Corentin {Dupont}},
  \bibinfo{person}{Raffaele {Giaffreda}}, {and} \bibinfo{person}{Luca
  {Capra}}.} \bibinfo{year}{2017}\natexlab{}.
\newblock \showarticletitle{Edge computing in IoT context: Horizontal and
  vertical Linux container migration}. In \bibinfo{booktitle}{\emph{Global
  Internet of Things Summit (GIoTS) 2017}}.
\newblock
\urldef\tempurl%
\url{https://doi.org/10.1109/GIOTS.2017.8016218}
\showDOI{\tempurl}


\bibitem[\protect\citeauthoryear{{Enes}, {Almeida}, {Baquero}, and
  {Leitão}}{{Enes} et~al\mbox{.}}{2019}]%
        {Enes2019}
\bibfield{author}{\bibinfo{person}{Vitor {Enes}}, \bibinfo{person}{Paulo~S.
  {Almeida}}, \bibinfo{person}{Carlos {Baquero}}, {and} \bibinfo{person}{João
  {Leitão}}.} \bibinfo{year}{2019}\natexlab{}.
\newblock \showarticletitle{Efficient Synchronization of State-Based CRDTs}. In
  \bibinfo{booktitle}{\emph{IEEE International Conference on Data Engineering
  (ICDE) 2019}}.
\newblock
\urldef\tempurl%
\url{https://doi.org/10.1109/ICDE.2019.00022}
\showDOI{\tempurl}


\bibitem[\protect\citeauthoryear{{Fox} and {Brewer}}{{Fox} and
  {Brewer}}{1999}]%
        {Fox1999}
\bibfield{author}{\bibinfo{person}{Armando {Fox}} {and}
  \bibinfo{person}{Eric~A. {Brewer}}.} \bibinfo{year}{1999}\natexlab{}.
\newblock \showarticletitle{Harvest, yield, and scalable tolerant systems}. In
  \bibinfo{booktitle}{\emph{Hot Topics in Operating Systems ({HotOS}) 1999}}.
\newblock
\urldef\tempurl%
\url{https://doi.org/10.1109/HOTOS.1999.798396}
\showDOI{\tempurl}


\bibitem[\protect\citeauthoryear{Hassas~Yeganeh and Ganjali}{Hassas~Yeganeh and
  Ganjali}{2012}]%
        {Feamstera}
\bibfield{author}{\bibinfo{person}{Soheil Hassas~Yeganeh} {and}
  \bibinfo{person}{Yashar Ganjali}.} \bibinfo{year}{2012}\natexlab{}.
\newblock \showarticletitle{Kandoo: A Framework for Efficient and Scalable
  Offloading of Control Applications}. In \bibinfo{booktitle}{\emph{Hot Topics
  in Software Defined Networks ({HotSDN}) 2012}}.
\newblock
\urldef\tempurl%
\url{https://doi.org/10.1145/2342441.2342446}
\showDOI{\tempurl}


\bibitem[\protect\citeauthoryear{{Jiménez} and {Schelén}}{{Jiménez} and
  {Schelén}}{2019}]%
        {Jimenez2019}
\bibfield{author}{\bibinfo{person}{Lara~L. {Jiménez}} {and}
  \bibinfo{person}{Olov {Schelén}}.} \bibinfo{year}{2019}\natexlab{}.
\newblock \showarticletitle{DOCMA: A Decentralized Orchestrator for
  Containerized Microservice Applications}. In \bibinfo{booktitle}{\emph{2019
  IEEE Cloud Summit}}.
\newblock
\urldef\tempurl%
\url{https://doi.org/10.1109/CloudSummit47114.2019.00014}
\showDOI{\tempurl}


\bibitem[\protect\citeauthoryear{{Kleppmann} and {Beresford}}{{Kleppmann} and
  {Beresford}}{2017}]%
        {Kleppmann2017}
\bibfield{author}{\bibinfo{person}{Martin {Kleppmann}} {and}
  \bibinfo{person}{Alastair~R. {Beresford}}.} \bibinfo{year}{2017}\natexlab{}.
\newblock \showarticletitle{A Conflict-Free Replicated JSON Datatype}.
\newblock \bibinfo{journal}{\emph{IEEE Transactions on Parallel and Distributed
  Systems}} \bibinfo{volume}{28}, \bibinfo{number}{10} (\bibinfo{year}{2017}),
  \bibinfo{pages}{2733--2746}.
\newblock
\urldef\tempurl%
\url{https://doi.org/10.1109/TPDS.2017.2697382}
\showDOI{\tempurl}


\bibitem[\protect\citeauthoryear{Kleppmann and Howard}{Kleppmann and
  Howard}{2020}]%
        {kleppmann2020byzantine}
\bibfield{author}{\bibinfo{person}{Martin Kleppmann} {and}
  \bibinfo{person}{Heidi Howard}.} \bibinfo{year}{2020}\natexlab{}.
\newblock \bibinfo{title}{Byzantine Eventual Consistency and the Fundamental
  Limits of Peer-to-Peer Databases}.
\newblock
\newblock
\showeprint[arxiv]{2012.00472}~[cs.DC]


\bibitem[\protect\citeauthoryear{Koponen, Casado, Gude, Stribling, Poutievski,
  Zhu, Ramanathan, Iwata, Inoue, Hama, and Shenker}{Koponen
  et~al\mbox{.}}{2010}]%
        {Koponen}
\bibfield{author}{\bibinfo{person}{Teemu Koponen}, \bibinfo{person}{Martin
  Casado}, \bibinfo{person}{Natasha Gude}, \bibinfo{person}{Jeremy Stribling},
  \bibinfo{person}{Leon Poutievski}, \bibinfo{person}{Min Zhu},
  \bibinfo{person}{Rajiv Ramanathan}, \bibinfo{person}{Yuichiro Iwata},
  \bibinfo{person}{Hiroaki Inoue}, \bibinfo{person}{Takayuki Hama}, {and}
  \bibinfo{person}{Scott Shenker}.} \bibinfo{year}{2010}\natexlab{}.
\newblock \showarticletitle{Onix: A Distributed Control Platform for
  Large-Scale Production Networks}. In \bibinfo{booktitle}{\emph{Operating
  Systems Design and Implementation ({OSDI}) 2010}}.
\newblock


\bibitem[\protect\citeauthoryear{Kr\'{o}l, Mastorakis, Oran, and
  Kutscher}{Kr\'{o}l et~al\mbox{.}}{2019}]%
        {Krol2019}
\bibfield{author}{\bibinfo{person}{Micha\l{} Kr\'{o}l},
  \bibinfo{person}{Spyridon Mastorakis}, \bibinfo{person}{David Oran}, {and}
  \bibinfo{person}{Dirk Kutscher}.} \bibinfo{year}{2019}\natexlab{}.
\newblock \showarticletitle{Compute First Networking: Distributed Computing
  Meets ICN}. In \bibinfo{booktitle}{\emph{Information-Centric Networking (ICN)
  2019}}.
\newblock
\urldef\tempurl%
\url{https://doi.org/10.1145/3357150.3357395}
\showDOI{\tempurl}


\bibitem[\protect\citeauthoryear{Kuenzer, Ivanov, Manco, Mendes, Volchkov,
  Schmidt, Yasukata, Honda, and Huici}{Kuenzer et~al\mbox{.}}{2017}]%
        {Kuenzer}
\bibfield{author}{\bibinfo{person}{Simon Kuenzer}, \bibinfo{person}{Anton
  Ivanov}, \bibinfo{person}{Filipe Manco}, \bibinfo{person}{Jose Mendes},
  \bibinfo{person}{Yuri Volchkov}, \bibinfo{person}{Florian Schmidt},
  \bibinfo{person}{Kenichi Yasukata}, \bibinfo{person}{Michio Honda}, {and}
  \bibinfo{person}{Felipe Huici}.} \bibinfo{year}{2017}\natexlab{}.
\newblock \showarticletitle{Unikernels Everywhere: The Case for Elastic CDNs}.
  In \bibinfo{booktitle}{\emph{Virtual Execution Environments (VEE) 2017}}.
\newblock
\urldef\tempurl%
\url{https://doi.org/10.1145/3050748.3050757}
\showDOI{\tempurl}


\bibitem[\protect\citeauthoryear{{Larsson}, {Gustafsson}, {Klein}, and
  {Elmroth}}{{Larsson} et~al\mbox{.}}{2020}]%
        {Larsson2020}
\bibfield{author}{\bibinfo{person}{Lars {Larsson}}, \bibinfo{person}{Harald
  {Gustafsson}}, \bibinfo{person}{Cristian {Klein}}, {and}
  \bibinfo{person}{Erik {Elmroth}}.} \bibinfo{year}{2020}\natexlab{}.
\newblock \showarticletitle{Decentralized Kubernetes Federation Control Plane}.
  In \bibinfo{booktitle}{\emph{Utility and Cloud Computing (UCC) 2020}}.
\newblock
\urldef\tempurl%
\url{https://doi.org/10.1109/UCC48980.2020.00056}
\showDOI{\tempurl}


\bibitem[\protect\citeauthoryear{Ongaro and Ousterhout}{Ongaro and
  Ousterhout}{2014}]%
        {Ongaro2019}
\bibfield{author}{\bibinfo{person}{Diego Ongaro} {and} \bibinfo{person}{John
  Ousterhout}.} \bibinfo{year}{2014}\natexlab{}.
\newblock \showarticletitle{In Search of an Understandable Consensus
  Algorithm}. In \bibinfo{booktitle}{\emph{{USENIX} Annual Technical Conference
  ({USENIX} {ATC}) 2014}}.
\newblock


\bibitem[\protect\citeauthoryear{Ren, Ananthanarayanan, Wierman, and Yu}{Ren
  et~al\mbox{.}}{2015}]%
        {Ren2015}
\bibfield{author}{\bibinfo{person}{Xiaoqi Ren}, \bibinfo{person}{Ganesh
  Ananthanarayanan}, \bibinfo{person}{Adam Wierman}, {and}
  \bibinfo{person}{Minlan Yu}.} \bibinfo{year}{2015}\natexlab{}.
\newblock \showarticletitle{Hopper: Decentralized Speculation-Aware Cluster
  Scheduling at Scale}. In \bibinfo{booktitle}{\emph{Special Interest Group on
  Data Communication (SIGCOMM) 2015}}.
\newblock
\urldef\tempurl%
\url{https://doi.org/10.1145/2785956.2787481}
\showDOI{\tempurl}


\bibitem[\protect\citeauthoryear{Rystsov}{Rystsov}{2018}]%
        {Rystsov2018}
\bibfield{author}{\bibinfo{person}{Denis Rystsov}.}
  \bibinfo{year}{2018}\natexlab{}.
\newblock \bibinfo{title}{CASPaxos: Replicated State Machines without logs}.
\newblock
\newblock
\showeprint[arxiv]{1802.07000}~[cs.DC]


\bibitem[\protect\citeauthoryear{{Sakic}, {Sardis}, {Guck}, and
  {Kellerer}}{{Sakic} et~al\mbox{.}}{2017}]%
        {Sakic2017}
\bibfield{author}{\bibinfo{person}{Ermin {Sakic}}, \bibinfo{person}{Fragkiskos
  {Sardis}}, \bibinfo{person}{Jochen~W. {Guck}}, {and}
  \bibinfo{person}{Wolfgang {Kellerer}}.} \bibinfo{year}{2017}\natexlab{}.
\newblock \showarticletitle{Towards adaptive state consistency in distributed
  SDN control plane}. In \bibinfo{booktitle}{\emph{IEEE International
  Conference on Communications (ICC) 2017}}.
\newblock
\urldef\tempurl%
\url{https://doi.org/10.1109/ICC.2017.7997164}
\showDOI{\tempurl}


\bibitem[\protect\citeauthoryear{Shapiro, Pregui{\c{c}}a, Baquero, and
  Zawirski}{Shapiro et~al\mbox{.}}{2011}]%
        {Preguica2018}
\bibfield{author}{\bibinfo{person}{Marc Shapiro}, \bibinfo{person}{Nuno
  Pregui{\c{c}}a}, \bibinfo{person}{Carlos Baquero}, {and}
  \bibinfo{person}{Marek Zawirski}.} \bibinfo{year}{2011}\natexlab{}.
\newblock \showarticletitle{Conflict-Free Replicated Data Types}. In
  \bibinfo{booktitle}{\emph{Stabilization, Safety, and Security of Distributed
  Systems}}.
\newblock


\bibitem[\protect\citeauthoryear{Stankovic}{Stankovic}{1984}]%
        {Stankovic}
\bibfield{author}{\bibinfo{person}{John~A Stankovic}.}
  \bibinfo{year}{1984}\natexlab{}.
\newblock \showarticletitle{Simulations of three adaptive, decentralized
  controlled, job scheduling algorithms}.
\newblock \bibinfo{journal}{\emph{Computer Networks (1976)}}
  \bibinfo{volume}{8}, \bibinfo{number}{3} (\bibinfo{year}{1984}),
  \bibinfo{pages}{199--217}.
\newblock
\showISSN{0376-5075}
\urldef\tempurl%
\url{https://doi.org/10.1016/0376-5075(84)90048-5}
\showDOI{\tempurl}


\bibitem[\protect\citeauthoryear{{Stankovic}}{{Stankovic}}{1985}]%
        {Stankovic1985}
\bibfield{author}{\bibinfo{person}{John~A. {Stankovic}}.}
  \bibinfo{year}{1985}\natexlab{}.
\newblock \showarticletitle{Stability and Distributed Scheduling Algorithms}.
\newblock \bibinfo{journal}{\emph{IEEE Transactions on Software Engineering}}
  \bibinfo{volume}{SE-11}, \bibinfo{number}{10} (\bibinfo{year}{1985}),
  \bibinfo{pages}{1141--1152}.
\newblock
\urldef\tempurl%
\url{https://doi.org/10.1109/TSE.1985.231862}
\showDOI{\tempurl}


\bibitem[\protect\citeauthoryear{van~der Linde, Leit\~{a}o, and
  Pregui\c{c}a}{van~der Linde et~al\mbox{.}}{2016}]%
        {VanderLinde}
\bibfield{author}{\bibinfo{person}{Albert van~der Linde},
  \bibinfo{person}{Jo\~{a}o Leit\~{a}o}, {and} \bibinfo{person}{Nuno
  Pregui\c{c}a}.} \bibinfo{year}{2016}\natexlab{}.
\newblock \showarticletitle{$\Delta$-CRDTs: Making $\delta$-CRDTs Delta-Based}.
  In \bibinfo{booktitle}{\emph{Principles and Practice of Consistency for
  Distributed Data ({PaPoC}) 2016}}.
\newblock
\urldef\tempurl%
\url{https://doi.org/10.1145/2911151.2911163}
\showDOI{\tempurl}


\bibitem[\protect\citeauthoryear{{Wu}, {Faleiro}, {Lin}, and
  {Hellerstein}}{{Wu} et~al\mbox{.}}{2018}]%
        {Anna}
\bibfield{author}{\bibinfo{person}{Chenggang {Wu}}, \bibinfo{person}{Jose
  {Faleiro}}, \bibinfo{person}{Yihan {Lin}}, {and} \bibinfo{person}{Joseph
  {Hellerstein}}.} \bibinfo{year}{2018}\natexlab{}.
\newblock \showarticletitle{Anna: A KVS for Any Scale}. In
  \bibinfo{booktitle}{\emph{IEEE 34th International Conference on Data
  Engineering (ICDE) 2018}}.
\newblock
\urldef\tempurl%
\url{https://doi.org/10.1109/ICDE.2018.00044}
\showDOI{\tempurl}


\bibitem[\protect\citeauthoryear{Zaharia, Borthakur, Sen~Sarma, Elmeleegy,
  Shenker, and Stoica}{Zaharia et~al\mbox{.}}{2010}]%
        {Zaharia2010}
\bibfield{author}{\bibinfo{person}{Matei Zaharia}, \bibinfo{person}{Dhruba
  Borthakur}, \bibinfo{person}{Joydeep Sen~Sarma}, \bibinfo{person}{Khaled
  Elmeleegy}, \bibinfo{person}{Scott Shenker}, {and} \bibinfo{person}{Ion
  Stoica}.} \bibinfo{year}{2010}\natexlab{}.
\newblock \showarticletitle{Delay Scheduling: A Simple Technique for Achieving
  Locality and Fairness in Cluster Scheduling}. In
  \bibinfo{booktitle}{\emph{European Conference on Computer Systems (EuroSys)
  2010}}.
\newblock
\urldef\tempurl%
\url{https://doi.org/10.1145/1755913.1755940}
\showDOI{\tempurl}


\end{thebibliography}

\end{document}